\documentclass{article}
\pdfoutput=1

\usepackage[final]{neurips_2021}

\usepackage{natbib}
\setcitestyle{square,numbers,comma}

\usepackage[pdftex]{graphicx}
\usepackage{algorithm}
\usepackage{algpseudocode}
\usepackage{multirow}

\usepackage[utf8]{inputenc} 
\usepackage[T1]{fontenc}    
\usepackage{hyperref}       
\usepackage{url}            
\usepackage{booktabs}       
\usepackage{amsfonts}       
\usepackage{nicefrac}       
\usepackage{microtype}      
\usepackage{xcolor}         

\title{An Empirical Evaluation of Zeroth-Order Optimization Methods on AI-driven Molecule Optimization}

\author{
    Elvin Lo \\
    Horace Greeley High School \\
    Chappaqua, NY 10514 \\
    \texttt{elvinlo922@gmail.com} \\
    \And
    Pin-Yu Chen \\
    IBM Research \\
    Yorktown Heights, NY 10598 \\
    \texttt{pin-yu.chen@ibm.com} \\
}

\begin{document}

\maketitle

\begin{abstract}
Molecule optimization is an important problem in chemical discovery and has been approached using many techniques, including generative modeling, reinforcement learning, genetic algorithms, and much more. Recent work has also applied zeroth-order (ZO) optimization, a subset of gradient-free optimization that solves problems similarly to gradient-based methods, for optimizing latent vector representations from an autoencoder. In this paper, we study the effectiveness of various ZO optimization methods for optimizing molecular objectives, which are characterized by variable smoothness, infrequent optima, and other challenges. We provide insights on the robustness of various ZO optimizers in this setting, show the advantages of ZO sign-based gradient descent (ZO-signGD), discuss how ZO optimization can be used practically in realistic discovery tasks, and demonstrate the potential effectiveness of ZO optimization methods on widely used benchmark tasks from the Guacamol suite. Code is available at: \textcolor{blue}{\url{https://github.com/IBM/QMO-bench}}.
\end{abstract}

\section{Introduction}

The goal of molecule optimization is to efficiently find molecules possessing desirable chemical properties. As the ability to effectively solve difficult molecule optimization tasks would greatly accelerate the discovery of promising drug candidates and decrease the immense resources necessary for drug development, significant efforts have been dedicated to designing molecule optimization algorithms leveraging a variety of techniques, including deep reinforcement learning \citep{olivecrona2017molecular}, genetic algorithms \citep{jensen2019graph}, Bayesian optimization \citep{tripp2021fresh}, variational autoencoders \citep{gomez2018automatic,jin2018junction}, and more. Several molecule optimization benchmark tasks have also been proposed, including similarity-based oracles \citep{brown2019guacamol} and docking scores \citep{garcia2022dockstring}. 

In this paper, we extend the work of \citet{hoffman2022optimizing}, who proposed the use of ZO optimization in their query-based molecule optimization (QMO) framework, an end-to-end framework which decouples molecule representation learning and property prediction. QMO iteratively optimizes a starting molecule, making it well suited for lead optimization tasks, but it can also start from random points and traverse large distances to find optimal molecules. In comparison with the work of \citet{hoffman2022optimizing} which experiments with only one optimizer, we experiment with variations of QMO using different ZO optimizers. Furthermore, we add more benchmark tasks from Guacamol \citep{brown2019guacamol} (whose use has been encouraged by the molecule optimization community \citep{tripp2021fresh, tripp2022evaluation} and used by \citet{gao2022sample} to benchmark many design algorithms in a standardized setting), and provide insights into the challenges of ZO optimization on molecular objectives.

\paragraph{Contributions} We evaluate several ZO optimization methods for the problem of molecule optimization in terms of convergence speed, convergence accuracy, and robustness to the unusual function landscapes (described further in Section \ref{sec:motivating}) of molecular objectives. We show that ZO-signGD \citep{liu2018signsgd} outperforms other methods for molecule optimization in not only speed but also accuracy, despite being known to have worse convergence accuracy for other problems like adversarial attacks \citep{liu2018signsgd}, which indicates that the sign operation is potentially more robust to the function landscapes of molecular objectives. Furthermore, we provide insights into practical application of ZO optimization in drug discovery scenarios for both lead optimization tasks and the discovery of novel molecules, as well as propose the use of a hybrid approach combining others models with QMO.

\paragraph{Related work} ZO optimization is a class of methods used for solving black-box problems by estimating gradients using only zeroth-order function evaluations and performing iterative updates as in first-order methods like gradient descent (GD) \citep{liu2020primer}. Many types of ZO optimization algorithms have been developed, including ZO gradient descent (ZO-GD) \citep{nesterov2017random}, ZO-signGD \citep{liu2018signsgd}, ZO adaptive momentum method (ZO-AdaMM, or ZO-Adam specifically for the Adam variant) \citep{chen2019zo}, and more \citep{lian2016comprehensive,li2022zeroth}. The optimality of ZO optimization methods has also been studied under given problem assumptions \citep{kornowski2021oracle}. ZO optimization methods have achieved impressive feats in adversarial machine learning, where they have been used for adversarial example generation in black-box settings and demonstrated comparable success to first-order white-box attacks \citep{chen2017zoo,tu2019autozoom}. They have also been shown to be able to generate contrastive explanations for black-box models \citep{dhurandhar2019model}. Finally, \citet{hoffman2022optimizing} showed how ZO optimization methods can also be applied to molecule optimization with their QMO framework.

\section{QMO: Background and Variations}

\subsection{QMO framework}
Following the QMO framework by \citet{hoffman2022optimizing}, we use an autoencoder to encode molecules with encoder $E:\mathcal{X}\mapsto \mathbb{R}^d$ and decode latent vectors with decoder $D:\mathbb{R}^d\mapsto\mathcal{X}$, where $\mathcal{X}$ denotes the discrete chemical space of all drug candidates. We denote with $\mathcal{O}:\mathcal{X}\mapsto\mathbb{R}$ a black-box oracle returning a scalar corresponding to a molecular property of interest (which may also be modified by adding losses related to other properties), and for ease of notation with the QMO framework, we define our optimization objective loss function as $f(\mathbf{z})=-\mathcal{O}(D(\mathbf{z}))$ for latent representations $\mathbf{z}\in\mathbb{R}^d$. As each function query $f(\mathbf{z})$ queries the oracle $\mathcal{O}$ with the decoded molecule corresponding to $\mathbf{z}$, one function query is equivalent to one oracle query.

In QMO, we use ZO optimization methods to navigate the latent space to solve $\mathrm{min}_{\mathbf{z}}f(\mathbf{z})$. Specifically, given a starting molecule and its latent representation $\mathbf{z}_0$, we iteratively update the current latent representation following some optimizer, as done in first-order gradient-based methods like gradient descent. But as we do not have any first-order oracle, we instead use gradients estimated using only evaluations of $f$ following some gradient estimator. The QMO framework, which closely follows a generic ZO optimization procedure, is summarized in Algorithm \ref{alg:qmo}.

\begin{algorithm}
    \caption{Generic QMO framework for molecule optimization}
    \label{alg:qmo}
    \begin{algorithmic}[1]
    \State \textbf{Inputs:} Starting molecule $x_0\in\mathcal{X}$, encoder $E$, decoder $D$, gradient estimation operation $\mathrm{estimate\_{gradient}}_f(\cdot)$ for function $f$, optimizer updating operation $\mathrm{update}(\cdot)$, number of iterations $T$, and learning rate $\alpha$
    \State $\mathcal{Z}_\mathrm{iterate} \gets \{\emptyset\}$
    \State $\mathbf{z}_0 \gets E(x_0)$
    \For{$t=0,1,...,T-1$}
        \State $\hat{\mathbf{g}}_t \gets \mathrm{estimate\_{gradient}}_f(\mathbf{z}_t, E, D)$
        \State $\mathbf{z}_{t+1} \gets \mathrm{update}(\mathbf{z}_t, \{\hat{\mathbf{g}}_i\}_{i=1}^{t}, \alpha, E, D)$
        \State $\mathcal{Z}_\mathrm{iterate} \gets \mathcal{Z}_\mathrm{iterate}\cup\{\mathbf{z}_{t+1}\}$
    \EndFor
    \end{algorithmic}
\end{algorithm}

In principle, QMO is a generic framework which can guide searches over any continuous learned representation based on any discrete space and use any ZO optimization method. \citet{hoffman2022optimizing} used the pre-trained SMILES-based \citep{weininger1988smiles} autoencoder (CDDD model) from \citet{winter2019learning} with embedding dimension $d=512$ and ZO-Adam. Here, we use the same autoencoder but consider several variations of QMO using different gradient estimators and optimizers to provide a comprehensive study on the effect of ZO optimization methods.

As a note, QMO is applicable to molecule optimization with design constraints. For example, given a set of property scores $\{p_i\}_{i=1}^I$ to be optimized with positive coefficients $\{\gamma_i\}_{i=1}^I$ and a set of property constraints $\{c_j\}_{j=1}^J$ with thresholds $\{\eta_j\}_{j=1}^J$, we can define the oracle as
\begin{equation}
\label{eq:constrained-mo}
\mathcal{O}(x)=\sum_{i=1}^{I}\gamma_i\cdot p_i(x)-\sum_{j=1}^{J}\mathrm{max}(\eta_j-c_j(x),0)
\end{equation}
where $x\in\mathcal{X}$. The vectors $\mathbf{z}\in\mathcal{Z}_\mathrm{iterate}$ not satisfying $c_j(D(\mathbf{z}))\geq\eta_j$ for all $j\in\{1,2,...,J\}$ can then be removed from $\mathcal{Z}_\mathrm{iterate}$.

\subsection{ZO gradient estimators}

We consider two main ZO gradient estimators. Both average finite differences over $Q$ independently sampled random perturbations $\{\mathbf{u}_q\}_{q=1}^{Q}$, include a smoothing parameter $\beta$, and follow the form: 
\begin{equation}
    \widehat{\nabla}f(\mathbf{z}) = \frac{\varphi(d)}{\beta\cdot Q}\sum_{q=1}^{Q}{[f(\mathbf{z}+\beta\mathbf{u}_q)-f(\mathbf{z})]\mathbf{u}_q}
\end{equation}
The two gradient estimators differ mainly on the sampling method for each random direction $\mathbf{u}_q$, and also by the dimension-dependent factor $\varphi(d)$. They are:

\begin{itemize}
    \item \textbf{Gaussian smoothing (GS)} \citep{nesterov2017random, flaxman2005online}: when we sample each direction from the uniform distribution $\mathcal{U}(\mathcal{S}(0,1))$ on the unit sphere. For GS, $\varphi(d)=d$.
    
    \item \textbf{Bernoulli smoothing-shrinkage (BeS-shrink)} \citep{gao2022generalizing}: when we craft each random direction by independently sampling each of its $d$ entries from $(B_{0.5}-0.5)/m$, where $B_{0.5}$ follows the Bernoulli distribution with probability $0.5$ and $m=\sqrt{\nicefrac{Q+d-1}{4Q}}$ is an optimal shrinking factor. For BeS-shrink, $\varphi(d)=1$.
\end{itemize}

By averaging over $Q$ random directions to decrease estimation error, the gradient estimation operation requires querying $Q+1$ different points (which are each decoded into a molecule and used to query oracle $\mathcal{O}$). We therefore require $Q+1$ oracle evaluations for each optimization iteration.

Additionally, because the above gradient estimators use a (forward) finite difference of 2 points to estimate the gradient for each random perturbation, we refer to it as a 2-point gradient estimator. An alternative to the 2-point GS and BeS-shrink gradient estimators are their 1-point alternatives, which instead have the form:
\begin{equation}
    \widehat{\nabla}f(\mathbf{z}) = \frac{\varphi(d)}{\beta\cdot Q}\sum_{q=1}^{Q}{f(\mathbf{z}+\beta\mathbf{u}_q)\mathbf{u}_q}
\end{equation}
Similar to 2-point gradient estimators, 1-point estimators require $Q+1$ oracle queries at each iteration (the estimation operation itself requires only $Q$ queries, but this does not account for querying the updated molecule after each iteration). However, 1-point estimators are not commonly used in practice due to higher variance.

\subsection{ZO optimizers}

We consider three main optimizers, each having its own updating operation that consists of computing a descent direction $\mathbf{m}_t$ and then updating the current point. Each optimizer can be paired with any ZO gradient estimator. The three are as follows:

\begin{itemize}
    \item \textbf{ZO gradient descent (ZO-GD)} \citep{nesterov2017random}: analogous to stochastic gradient descent (SGD) in the first-order stochastic setting. ZO-GD uses the current gradient estimate as the descent direction $\mathbf{m}_t = \hat{\mathbf{g}}_t$ and updates the current point via the rule $\mathbf{z}_{t+1} = \mathbf{z}_t - \alpha\mathbf{m}_t$.
    
    \item \textbf{ZO sign-based gradient descent (ZO-signGD)} \citep{liu2018signsgd}: analogous to sign-based SGD (signSGD) \citep{bernstein2018signsgd} in the first-order stochastic setting. ZO-signGD uses the same point updating rule as ZO-GD but instead uses the sign of the current estimate as the descent direction $\mathbf{m}_t = \mathrm{sign}(\hat{\mathbf{g}}_t)$, where $\mathrm{sign}(\cdot)$ denotes the element-wise sign operation.
    
    \item \textbf{ZO-Adam} \citep{chen2019zo}: analogous to Adam \citep{kingma2015adam} in the first-order stochastic setting. ZO-Adam adopts a momentum-type descent direction and an adaptive learning rate.
\end{itemize}

\begin{table}
    \caption{Summary of ZO optimization methods considered.}
    \label{tab:zooms}
    \centering
    \begin{tabular}{l|l|c}
        \toprule
        ZO optimization method & Gradient estimator & Optimizer \\
        \midrule
        Adam-2p-BeS-shrink & 2-point BeS-shrink & Adam \\
        Adam-2p-GS & 2-point GS & Adam \\
        GD-2p-BeS-shrink & 2-point BeS-shrink & GD \\
        GD-2p-GS & 2-point GS & GD \\
        signGD-2p-BeS-shrink & 2-point BeS-shrink & signGD \\
        signGD-2p-GS & 2-point GS & signGD \\
        \bottomrule
    \end{tabular}
\end{table}

The ZO optimization methods compared in this paper are summarized in Table \ref{tab:zooms}.

\subsection{Motivating the comparison of ZO optimization methods for molecule optimization}
\label{sec:motivating}

We motivate our comparison of optimizers not only in terms of convergence speed and convergence accuracy, but also in terms of robustness to the unfriendly function landscapes of molecular objectives. Indeed, molecule optimization is made difficult by variable function smoothness due to "activity cliffs" in the molecular space where small structural changes cause large changes in oracle values \citep{tripp2022evaluation}. As optima are infrequent, there are also large and extremely "flat" unfavorable regions in the space, where the oracle values change minimally and may be very small. Furthermore, because our objective function $f$ obtains values by querying the oracle $\mathcal{O}$ using discrete molecular representations obtained from decoding the latent vectors, the function landscape is made discrete and thus further non-smooth (i.e., the function value may have a discrete "jump" at the borders between adjacent regions of latent vectors which decode to different molecules, see Fig. \ref{fig:landscapes}). Thus, being able to effectively navigate the latent chemical space and not get stuck in unfavorable regions is an important and non-trivial attribute to pursue in optimization methods.

Sign-based gradient descent is known to be effective in achieving fast convergence speed in stochastic settings: in the stochastic first-order oracle setting, \citet{bernstein2018signsgd} showed that signSGD could have faster empirical convergence speed than SGD, and in the zeroth-order stochastic setting \citet{liu2018signsgd} similarly showed that ZO-signSGD has faster convergence speed than many ZO optimization methods at the cost of worse accuracy (i.e., converging only to the neighborhood of an optima). The fast convergence of sign-based methods is motivated by the idea that the sign operation is more robust to stochastic noise, and though our formulation of molecule optimization is non-stochastic, the sign operation is potentially more robust to the difficult landscapes of molecular objective functions. Adaptive momentum methods like Adam also make use of the sign of stochastic gradients for determining the descent direction in addition to variance adaption \citep{balles2018dissecting}, and thus ZO-Adam may also show robustness to the function landscape.

\section{Practical Usage of QMO for Drug Discovery} 

We imagine QMO can be applied for two main cases: 1) identifying novel lead molecules (finding molecules significantly different from known leads), and 2) lead optimization (finding slightly modified versions of known leads).

For the former application case, it may be counterproductive to use known leads as the starting molecule in QMO, as those leads may be in the close neighborhood of a local optima (or a local optima themselves) in the function landscape, in which the optimizer would likely get stuck (preventing the exploration of different areas of the latent chemical space). Instead, it may be more promising to start at a random point in the chemical space. QMO also has the advantage that it guides search without the use of a training set, which aids in finding candidates vastly different from known molecules. However, finding a highly diverse set of novel leads may be unlikely within a single run of QMO as the optimization methods converge to some neighborhood, meaning that multiple random restarts would likely be necessary to discover a diverse set of lead molecules.

For the latter application case, it is much more sensible to use known leads as the starting molecule input to QMO. Additionally, rather than using an oracle $\mathcal{O}$ evaluating only the main desired drug property (i.e., activity against a biological target), it may be advantageous to use a modified oracle. For example, \citet{hoffman2022optimizing} apply QMO for lead optimization of known SARS-CoV-2 main protease inhibitors and antimicrobial peptides (AMPs) following the constrained molecule optimization setting of (\ref{eq:constrained-mo}), with pre-trained property predictors for each task. They set similarity to the original lead molecule as the property score $p_\mathrm{sim}$ to be optimized, and set constraints on properties of interest (binding affinity $c_\mathrm{aff}$ for the SARS-CoV-2 task, or toxicity prediction value $c_\mathrm{tox}$ and AMP prediction value $c_\mathrm{AMP}$ for the AMP task). In these formulations, the main optimization objective is actually molecular similarity rather than the main properties of interest.

\paragraph{Hybrid optimization: Integrating QMO with generative models} Additionally, in this paper we propose to integrate QMO with other models in a hybrid approach: namely, we can use molecules generated by other models as input to QMO, which will then iteratively optimize each of the inputted starting molecules. By using other models to generate good lead molecules close to optima, we can then use QMO to provide a more refined search that may incorporate additional design constraints. Overall, a hybrid approach could be a query-efficient way to generate drug candidates satisfying multiple design constraints.

\section{Experiments}

To benchmark QMO, we select three tasks (oracles) from the popular benchmarking suite Guacamol \citep{brown2019guacamol}: perindopril\_mpo (finding molecules similar to perindopril but with a different number of aromatic rings), zaleplon\_mpo (finding molecules similar to zaleplon but a different molecular formula), and deco\_hop (maximizing similarity to a SMILES string while keeping or excluding specific SMARTS patterns). These represent three of the main categories of Guacamol oracles: similarity-based multi-property objectives, isomer-based objectives, and SMARTS-based objectives, respectively. The zaleplon\_mpo task is known to be particularly difficult \citep{tripp2022evaluation}. We also select two baselines, graph-based genetic algorithm (Graph-GA) \citep{jensen2019graph} and Gaussian process Bayesian optimization (GPBO) \citep{tripp2021fresh, srinivas2010gaussian}, both of which are known to be high-performing molecule optimization algorithms \citep{gao2022sample}.

For each of these tasks, we run experiments using QMO only, baselines only, and hybrid approaches.
\begin{itemize}
    \item When running experiments using QMO only, we run with several different ZO optimization methods and try $Q=\{30,50,100\}$ for each, set $T=1000$ iterations for perindopril\_mpo and zaleplon\_mpo or $T=200$ for deco\_hop, and average runs from 20 distinct starting molecules with 2 random restarts each (40 runs total). For each task, we choose to use the 20 lowest-scoring molecules on the oracle from the ZINC 250K dataset \citep{sterling2015zinc} as the starting molecules. We do this in order to show that QMO can find solutions even when starting far from any high scoring molecules, which we would likely need to do when searching for novel lead molecules. 
    \item When running baseline models alone, we average runs with two random seeds and limit the number of oracle queries to 10K. 
    \item When running hybrid approaches, for each baseline model we use a portion of the 10K query budget to run the model (4K queries for Graph-GA and 2K for GPBO) and use the remaining query budget to optimize only the top generated molecule using QMO with the ZO-signGD optimizer and 2-point GS gradient estimator (QMO-sign-2p-GS) with $Q=49$.
\end{itemize}

Note that for these experiments, we consider only the score of the top 1 scoring molecule found so far for a given run. Additionally, we run QMO only with 2-point gradient estimators, though we also compare 1-point estimators for QED \citep{bickerton2012quantifying} optimization in Appendix \ref{app:1p} where we verify the advantage of 2-point estimators.

\subsection{Function landscapes of selected Guacamol objectives}

\begin{figure}
    \centering
    \includegraphics[width=\linewidth]{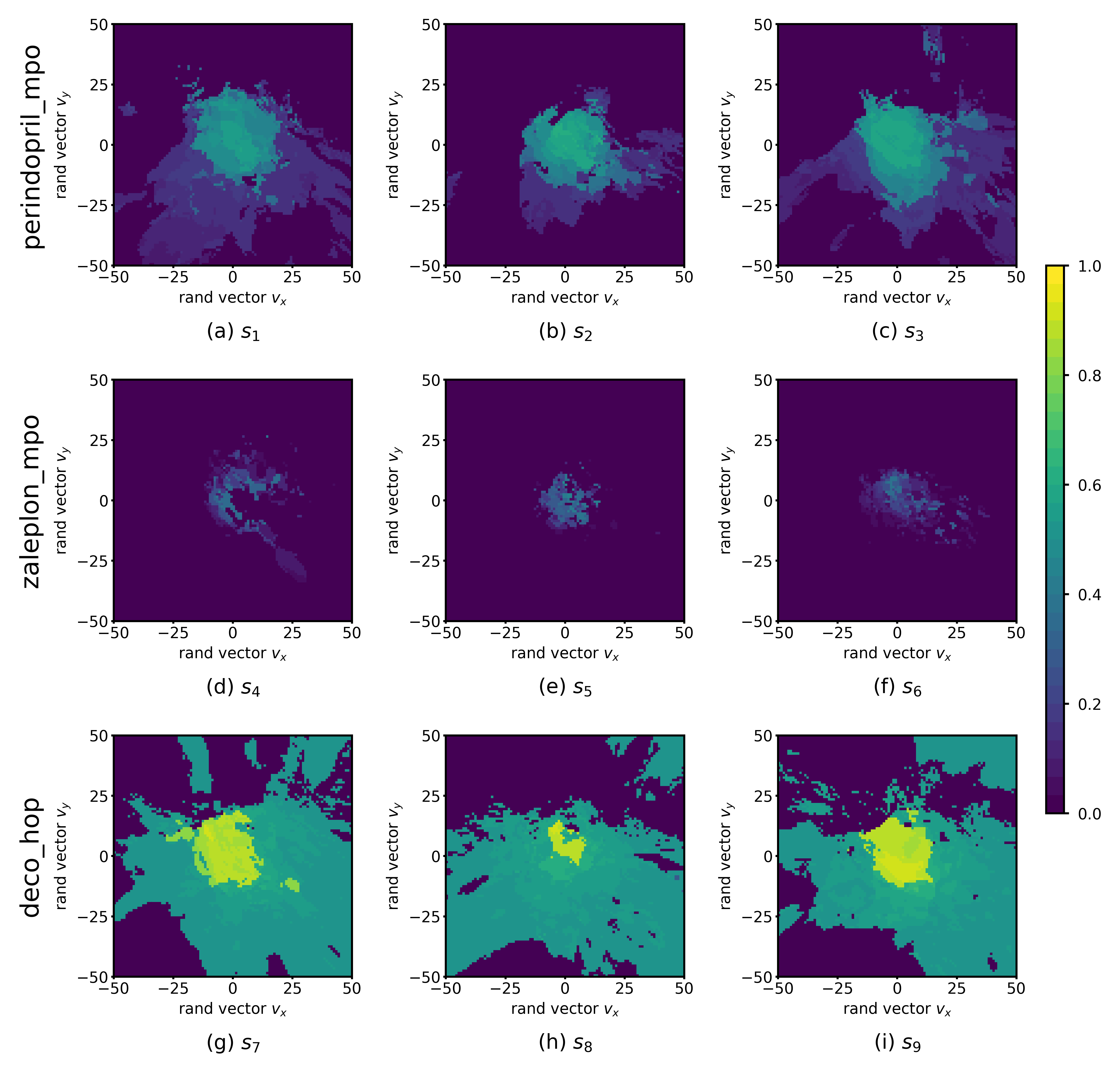}
    \caption{Function landscapes for various optimized molecules found by QMO. SMILES strings are listed in Appendix \ref{app:landscapes-smiles}.}
    \label{fig:landscapes}
\end{figure}

Fig. \ref{fig:landscapes} shows the function landscapes of the selected Guacamol objectives. The origin corresponds to an optimized latent vector found by QMO, and the vector is perturbed along two random directions $v_x$ and $v_y$ sampled from the uniform distribution on the unit sphere.

As shown, the zaleplon\_mpo task has the smallest central area consisting of high scoring molecules and a relatively flat landscape elsewhere, meaning that the QMO optimizer needs to traverse a very flat unfavorable region to enter a very small optimal neighborhood. This matches the observation that zaleplon\_mpo is a highly difficult task. The deco\_hop task, while not nearly as difficult of a task, still exhibits a very discrete jump in values around the central region, which makes it more difficult for the QMO optimizer to find the true optimal neighborhood. Finally, perindopril\_mpo appears to be the most smooth function. The optimal central area is larger than for zaleplon\_mpo, and the discrete jumps in function values are not as large as in the other tasks.

\subsection{Convergence of ZO optimization methods}
\label{sec:convergence}

\begin{figure}
    \centering
    \includegraphics[width=\linewidth]{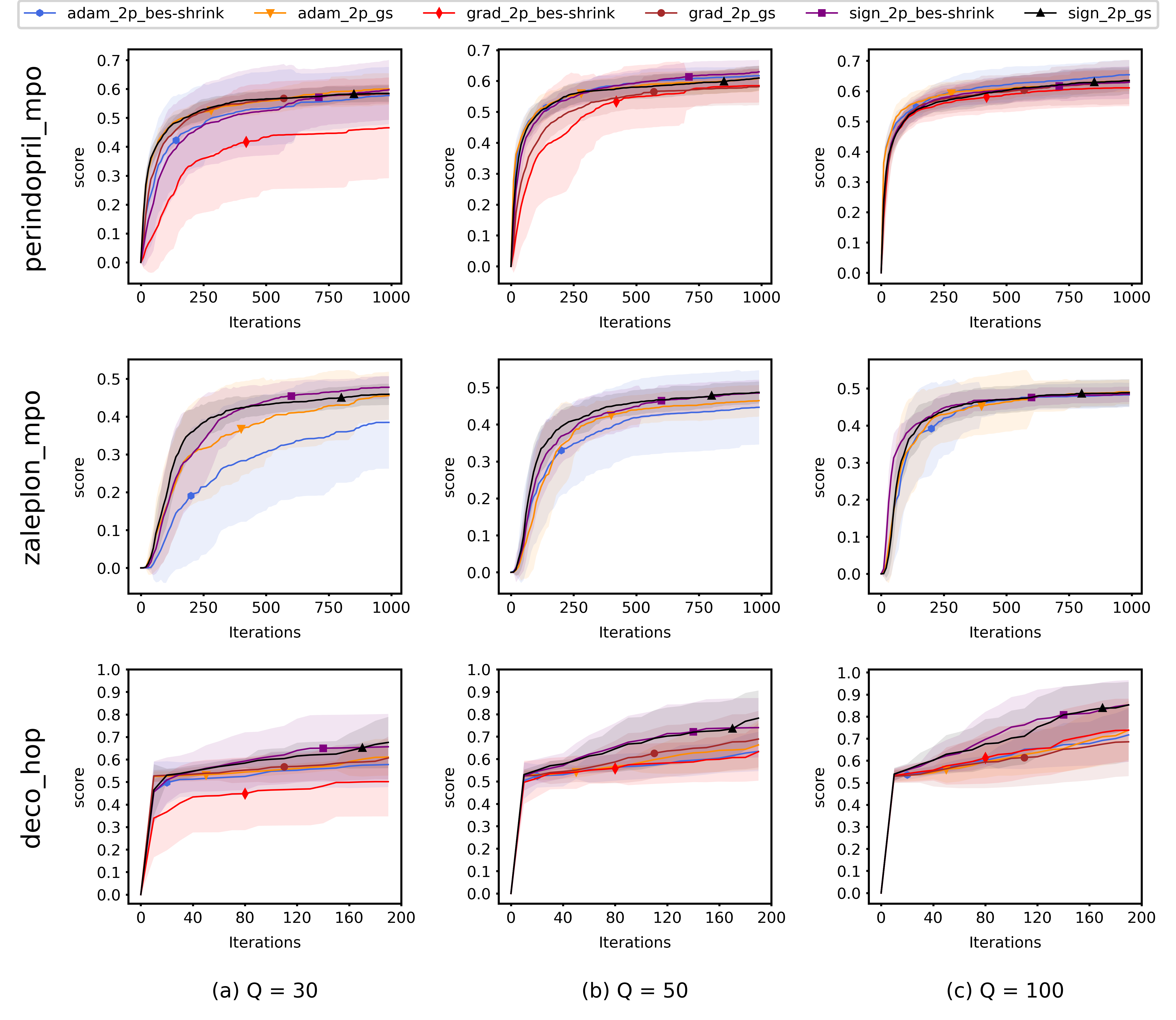}
    \caption{Convergence of QMO with different ZO optimizers.}
    \label{fig:convergence}
\end{figure}

Fig. \ref{fig:convergence} shows the results from experiments run using QMO only and compares the convergence of ZO optimization methods with different $Q$. Here, adam\_2p\_bes-shrink refers to QMO using the ZO-Adam optimizer with the 2-point BeS-shrink gradient estimator (QMO-Adam-2p-BeS-shrink), and similarly for the other ZO optimization methods. Diversity scores of the molecules found by QMO are reported in Appendix \ref{app:diversity}.

Overall, the results indicate that ZO-signGD is not only the most query-efficient method, but also the most robust to difficult function landscapes of molecular objectives. Compared with ZO-Adam, ZO-signGD converges with both higher speed and better accuracy for most settings, and the difference is especially clear for low $Q$ and less smooth functions like zaleplon\_mpo and deco\_hop. The improved convergence accuracy of ZO-signGD compared to ZO-Adam is particularly interesting as ZO-Adam converges with much greater accuracy in other problems like adversarial example generation \citep{chen2019zo}, thus showing the challenges presented by molecular objectives and the improved robustness of ZO-signGD to their function landscapes.

Finally, as a note, both ZO-Adam and ZO-signGD outperform ZO-GD. In fact, ZO-GD is completely unsuccessful for the zaleplon\_mpo task: even when searching a wide range of hyperparameters and testing several molecules, ZO-GD is unable to find any molecules with zaleplon\_mpo scores above 0.2 within the first 100 iterations, and often cannot even get above 0.01. Inspection revealed that the gradient vectors were too small for ZO-GD to make meaningful point updates. Thus, full zaleplon\_mpo experiments were not run using ZO-GD. In addition, results from GS and BeS-shrink gradient estimators do not differ greatly, though GS seems to converge faster with lower $Q$.

\subsection{Query efficiency of QMO versus other approaches}
\label{sec:queryefficiency}

\begin{figure}
    \centering
    \includegraphics[width=\linewidth]{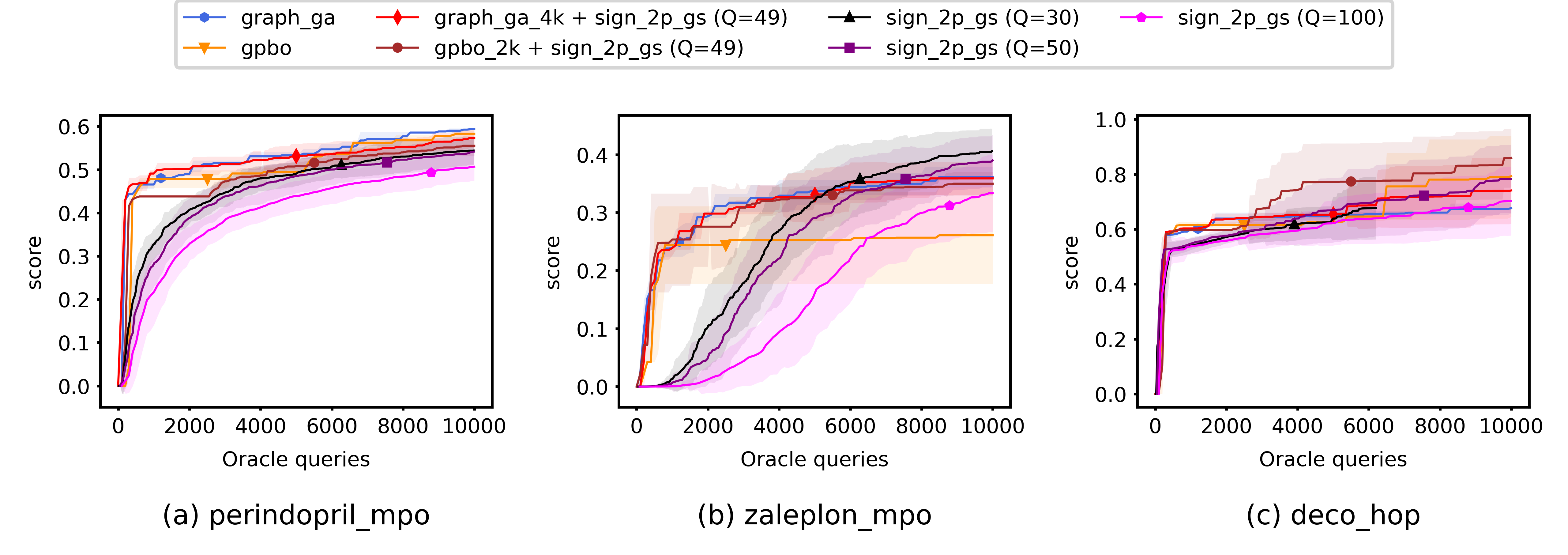}
    \caption{Optimization curves of QMO, generative models, and hybrid methods.}
    \label{fig:queryefficiency}
\end{figure}

Fig. \ref{fig:queryefficiency} shows the optimization curves when limiting optimization to a query 10K budget, including experiments run using QMO only (specifically, only QMO-sign-2p-GS is shown), baseline models only, and hybrid approaches. Precise numbers and area under curve (AUC) scores are also reported in Appendix \ref{tab:curve-scores}.

The baseline models (Graph-GA and GPBO) demonstrate faster convergence speed than QMO alone, and the relative convergence accuracies of all methods differ slightly for each task and can be said to be comparable overall. However, the hybrid approaches combining baseline models with QMO (i.e., Graph GA + sign\_2p\_gs) produce similar curves to their baseline model counterparts even for zaleplon\_mpo and deco\_hop, where QMO has higher convergence accuracy than the baseline models, so further investigation may be necessary to effectively integrate QMO into hybrid approaches.

\section{Conclusion}

In this paper, we study the application of ZO optimization methods to molecule optimization. Through experimentation on tasks from the Guacamol suite, we show that ZO-signGD outperforms ZO-Adam and ZO-GD, especially for more difficult function landscapes with small regions of optima, flat regions, and discrete jumps. Accordingly, we observe that the sign operation can increase robustness to the difficult function landscapes of molecular objectives, while also achieving higher query efficiency compared to other optimizer updating methods. We also discuss how the generic QMO framework can be applied practically in realistic drug discovery scenarios, which includes a hybrid approach with other models.

To conclude, we would like to mention a few limitations of this study. Synthesizability of molecules is not accounted for, though one possible approach is to modify the objective function with a synthesizability loss. Additionally, the effect of autoencoder choice and latent dimension are not thoroughly investigated for the selected benchmark tasks, though \citet{hoffman2022optimizing} provide analysis for their antimicrobial peptides task. Finally, while \citet{hoffman2022optimizing} also show that training an oracle prediction model (to predict scores based on latent representations) has significant disadvantages in optimization accuracy compared to always using the oracle itself, we do not thoroughly investigate the impact it would have on the objective function landscapes in the latent space.


\bibliography{refs}

\newpage

\section*{Appendix}

\appendix
\counterwithin{figure}{section}
\counterwithin{table}{section}
\renewcommand\thefigure{\thesection\arabic{figure}}
\renewcommand\thetable{\thesection\arabic{table}}

\section{Implementation details}
\label{app:implementation}

\citet{winter2019learning} showed that their CDDD autoencoder model has a high validity rate of 97\%, even when traversing a large distance from the valid latent representations of randomly picked molecules. In our implementation of QMO, we dealt with decode failures by assigning a penalty score of 0.1 less than the score of the starting molecule, $f(\mathbf{z}_0)-0.1$.

Also, we only considered the molecules generated after each optimization iteration. That is, we did not consider the $Q$ molecules obtained from decoding the perturbed latent vectors $\{\mathbf{z}+\beta\mathbf{u}_q\}_{q=1}^Q$ (used for estimating gradients) in $\mathcal{Z}_\mathrm{iterate}$ despite that they were also used to query the oracle $\mathcal{O}$. Especially in a realistic drug discovery scenario where oracle evaluations are highly expensive, we would of course want to also consider these molecules in case they exhibit good properties. In addition, while we considered there to be $Q+1$ oracle evaluations necessary for each optimization iteration, the actual amount would be lower in practice as some of the perturbed latent vectors would decode to the same molecule since the perturbations are small (and a small number of latent vectors would also decode to no valid input).

All experiments were run using Google Colab, and code for QMO is available at: \textcolor{blue}{\url{https://github.com/IBM/QMO-bench}}. For the Graph-GA and GPBO baseline models, we adopt the implementation of \citet{gao2022sample}.

\section{Additional results}
\label{app:additionalresults}

\subsection{Comparing 1-point and 2-point gradient estimators}
\label{app:1p}

\begin{figure}[b]
    \centering
    \includegraphics[width=\linewidth]{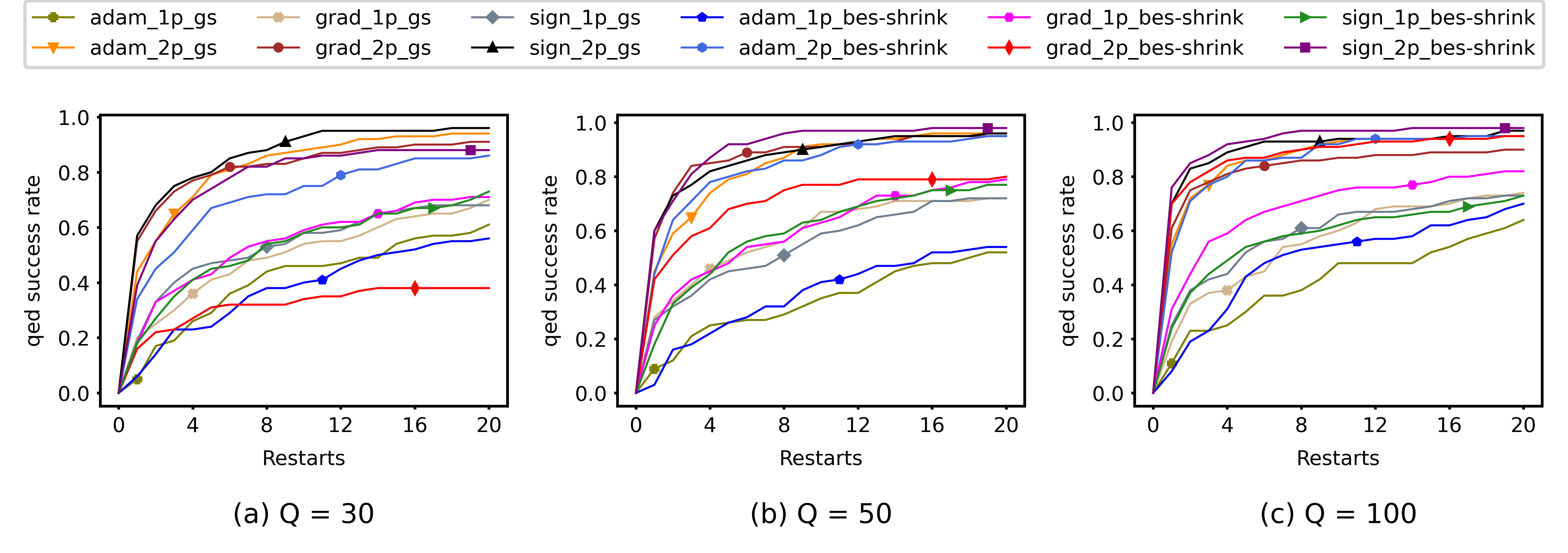}
    \caption{Optimization of QED with different ZO optimizers.}
    \label{fig:qedresults}
\end{figure}

Though we ran only 2-point gradient estimators on the Guacamol tasks, we also compared 1-point gradient estimators on the QED \citep{bickerton2012quantifying} objective. Specifically, following the setup of \citet{hoffman2022optimizing}, we defined a minimum similarity threshold of 0.4 (and did not consider molecules with similarity less than 0.4 to the starting molecule) and set the oracle $\mathcal{O}(x)=4\cdot\mathrm{QED}(x)-\mathrm{max}(0.4-\mathrm{sim}(x, x_0), 0)$ for molecule $x\in\mathcal{X}$ and starting molecule $x_0$, where $\mathrm{sim(\cdot)}$ denotes Tanimoto similarity with Morgan fingerprints. We selected 100 molecules with QED scores in $[0.7,0.8]$ from the test set in \citep{jin2018learning} (who extracted the molecules from ZINC) and optimized each with $T=20$ iterations and 20 random restarts each. We consider an optimized molecule a success if its QED scores falls in $[0.9,1.0]$, and we visualize in Fig. \ref{fig:qedresults} how many random restarts are necessary for different ZO optimization methods to achieve a given success rate. As shown, 2-point gradient estimators achieve significantly higher success rates than their 1-point counterparts given the same number of random restarts.

\subsection{SMILES strings of molecules found by QMO}
\label{app:landscapes-smiles}

The SMILES strings from Fig. \ref{fig:landscapes} are as follows: 
\begin{itemize}
    \item $s_1$ = \protect\nolinkurl{CCCC(NC(C)Cn1c(C2CCCCC2)nc2cc(C(=O)O)ccc21)C(=O)OCC}
    \item $s_2$ = \protect\nolinkurl{CCCC(C(=O)OCC)c1nc2cc(C(=O)O)ccc2n1C1CCCCC1C(C)=O}
    \item $s_3$ = \protect\nolinkurl{CCCC(C)n1c(C(=O)NC(C)C(=O)OCC)cc2cc(C3CCCCC3)cc(C(=O)O)c21}
    \item $s_4$ = \protect\nolinkurl{COc1cc(C(=O)NCC23CC=CC2C3)nc2c(C\#N)cccc12}
    \item $s_5$ = \protect\nolinkurl{CCC1(CC)C(C(=O)Nc2cccc(F)c2)=CN=C2C=C(C\#N)C(=O)N21}
    \item $s_6$ = \protect\nolinkurl{CCC12CC(CO1)N(C(=O)c1cnc3cccc(C\#N)c(=O)c3c1)C2}
    \item $s_7$ = \protect\nolinkurl{COc1cc2ncnc(Nc3ccc(F)c4ncccc34)c2cc1C(C)C}
    \item $s_8$ = \protect\nolinkurl{CCCCOc1ncccc1C(=O)CNc1ncnc2cc(OC(F)F)c(F)cc12}
    \item $s_9$ = \protect\nolinkurl{COc1cc(Nc2ncnc3cc(OC)c([N+](=O)[O-])cc23)ccc1F}.
\end{itemize}

\subsection{Diversity metrics}
\label{app:diversity}

Table \ref{tab:diversity} shows the diversity of the QMO optimized molecules from Section \ref{sec:convergence} (optimized using sign\_2p\_gs with $Q=50$). For each starting molecule in the test set, the best molecule found after two random restarts was used, showing the diversity of molecules that can be generated from different starting points. \citet{hoffman2022optimizing} also showed how different random restarts starting from the same starting point can find diverse candidates.

\begin{table}[ht]
    \caption{Diversity of 20 QMO optimized molecules from different starting points.}
    \label{tab:diversity}
    \centering
    \begin{tabular}{l|c|c}
        \toprule
        Task & Average score & Diversity \\
        \midrule
        perindopril\_mpo & 0.628 & 0.678 \\
        zaleplon\_mpo & 0.500 & 0.805 \\
        deco\_hop & 0.859 & 0.664 \\
        \bottomrule
    \end{tabular}
\end{table}

\subsection{Query efficiency tables}
\label{tab:curve-scores}

Scores from Fig. \ref{fig:queryefficiency} are summarized below in Tables \ref{tab:auc-perindopril}, \ref{tab:auc-zaleplon}, and \ref{tab:auc-deco}.

\begin{table}
    \caption{Scores for perindopril\_mpo with various query budgets.}
    \label{tab:auc-perindopril}
    \centering
    \begin{tabular}{l|c|c|c|c|c|c}
        \toprule
        Methods & AUC & 500 q & 1000 q & 2000 q & 5000 q & 10000 q \\
        \midrule
        graph\_ga & \textbf{0.527} & 0.453 & 0.465 & 0.490 & \textbf{0.533} & \textbf{0.593} \\
        gpbo & 0.502 & 0.446 & 0.478 & 0.478 & 0.494 & 0.583 \\
        sign\_2p\_gs ($Q=30$) & 0.456 & 0.219 & 0.327 & 0.408 & 0.490 & 0.544 \\
        sign\_2p\_gs ($Q=50$) & 0.441 & 0.176 & 0.284 & 0.388 & 0.485 & 0.541 \\
        sign\_2p\_gs ($Q=100$) & 0.395 & 0.101 & 0.218 & 0.330 & 0.439 & 0.507 \\
        graph\_ga\_4k + sign\_2p\_gs ($Q=49$) & 0.522 & \textbf{0.466} & \textbf{0.491} & \textbf{0.501} & 0.531 & 0.572 \\
        gpbo\_2k + sign\_2p\_gs ($Q=49$) & 0.487 & 0.435 & 0.438 & 0.438 & 0.508 & 0.555 \\
        \bottomrule
    \end{tabular}
\end{table}

\begin{table}
    \caption{Scores for zaleplon\_mpo with various query budgets.}
    \label{tab:auc-zaleplon}
    \centering
    \begin{tabular}{l|c|c|c|c|c|c}
        \toprule
        Methods & AUC & 500 q & 1000 q & 2000 q & 5000 q & 10000 q \\
        \midrule
        graph\_ga & \textbf{0.315} & 0.167 & 0.239 & 0.294 & \textbf{0.337} & 0.362 \\
        gpbo & 0.241 & 0.172 & 0.244 & 0.244 & 0.253 & 0.261 \\
        sign\_2p\_gs ($Q=30$) & 0.259 & 0.001 & 0.013 & 0.105 & 0.321 & \textbf{0.406} \\
        sign\_2p\_gs ($Q=50$) & 0.233 & 0.000 & 0.007 & 0.048 & 0.291 & 0.390 \\
        sign\_2p\_gs ($Q=100$) & 0.158 & 0.000 & 0.000 & 0.013 & 0.168 & 0.333 \\
        graph\_ga\_4k + sign\_2p\_gs ($Q=49$) & 0.314 & 0.183 & 0.239 & \textbf{0.298} & 0.331 & 0.359 \\
        gpbo\_2k + sign\_2p\_gs ($Q=49$) & 0.307 & \textbf{0.223} & \textbf{0.254} & 0.276 & 0.329 & 0.350 \\
        \bottomrule
    \end{tabular}
\end{table}

\begin{table}
    \caption{Scores for deco\_hop with various query budgets.}
    \label{tab:auc-deco}
    \centering
    \begin{tabular}{l|c|c|c|c|c|c}
        \toprule
        Methods & AUC & 500 q & 1000 q & 2000 q & 5000 q & 10000 q \\
        \midrule
        graph\_ga & 0.634 & 0.580 & 0.600 & \textbf{0.638} & 0.650 & 0.676 \\
        gpbo & 0.663 & 0.587 & \textbf{0.615} & 0.615 & 0.626 & 0.792 \\
        sign\_2p\_gs ($Q=30$) & 0.582 & 0.529 & 0.539 & 0.572 & 0.638 & - \\
        sign\_2p\_gs ($Q=50$) & 0.652 & 0.529 & 0.548 & 0.576 & 0.669 & 0.783 \\
        sign\_2p\_gs ($Q=100$) & 0.604 & 0.522 & 0.537 & 0.558 & 0.622 & 0.702 \\
        graph\_ga\_4k + sign\_2p\_gs ($Q=49$) & 0.661 & \textbf{0.592} & 0.602 & 0.637 & 0.655 & 0.741 \\
        gpbo\_2k + sign\_2p\_gs ($Q=49$) & \textbf{0.716} & 0.591 & 0.597 & 0.597 & \textbf{0.772} & \textbf{0.859} \\
        \bottomrule
    \end{tabular}
\end{table}

\section{Other ZO optimization methods}
Other than the ZO optimization methods considered here, ZO stochastic coordinate descent (ZO-SCD) \citep{lian2016comprehensive} was also tested. However, because the coordinate-wise gradient estimator relies on perturbing coordinates individually, the perturbed vector embeddings used to estimate gradients almost always decoded back to the same molecule as the original non-perturbed vector. In other words, the autoencoder almost always perceived the embedding vectors with all coordinates the same but one coordinate to be the same molecule, so the coordinate-wise gradient estimates were almost always zero.

\section{Hyperparameter tuning}
\label{hyperparams}

Aside from the number of random perturbations $Q$, there are two other main hyperparameters for each of the ZO optimization methods compared: function smoothing parameter $\beta$, and learning rate $\alpha$. The value $\beta=10$ was used for all tasks as it was found to work well with the CDDD model. Consistent with \citet{hoffman2022optimizing}, we find that $\beta=1$ or below does not work well (as gradients cannot be accurately approximated without sufficient smoothing) and $\beta=100$ or above results in many decode failures. When trying a few molecules with $\beta$ values between this range (including $\beta=\{5,10,20,50\}$ for each task, $\beta=10$ still performed best for the majority of molecules. The tuning of $\alpha$ is shown in Table \ref{tab:tuning} and Table \ref{tab:tuning-qed}. As a note, $\alpha$ larger than the largest tested values for each optimization method often resulted in many decode failures, so even if the best $\alpha$ was the largest value tested, choosing notably larger $\alpha$ (greater by more than a factor of 2) may not be a good idea. Also, for ZO-Adam, two additional hyperparameters are used for the adaptive learning rate: the exponential averaging parameters $\beta_1$ and $\beta_2$. For these parameters, we use the default values used by the PyTorch Adam implementation, $\beta_1=0.9$ and $\beta_2=0.999$.

\begin{table}[ht]
    \caption{Tuning of learning rate $\alpha$ for Guacamol tasks. Scores correspond to the average scores after optimizing 20 molecules with 2 random restarts each (40 trials total) for $T=1000$ iterations.}
    \label{tab:tuning}
    \centering
    \begin{tabular}{l|l|c|c|c|c}
        \toprule
        Task & Methods & Learning rate $\alpha$ & $Q=30$ & $Q=50$ & $Q=100$ \\
        \midrule
        \multirow{16}{*}{perindopril\_mpo} & \multirow{3}{*}{adam\_2p\_bes-shrink} & 0.1 & 0.555 & 0.598 & 0.607 \\
        & & 0.2 & \textbf{0.578} & 0.617 & 0.635 \\
        & & 0.3 & 0.564 & \textbf{0.617} & \textbf{0.654} \\
        \cmidrule{2-6}
        & \multirow{3}{*}{adam\_2p\_gs} & 0.1 & \textbf{0.600} & 0.600 & 0.604 \\
        & & 0.2 & 0.589 & \textbf{0.611} & 0.635 \\
        & & 0.3 & 0.560 & 0.605 & \textbf{0.637} \\
        \cmidrule{2-6}
        & \multirow{2}{*}{grad\_2p\_bes-shrink} & 30.0 & 0.429 & \textbf{0.585} & \textbf{0.611} \\
        & & 50.0 & \textbf{0.466} & 0.555 & 0.598 \\
        \cmidrule{2-6}
        & \multirow{2}{*}{grad\_2p\_gs} & 2.0 & \textbf{0.584} & \textbf{0.582} & 0.571 \\
        & & 5.0 & 0.500 & 0.566 & \textbf{0.630} \\
        \cmidrule{2-6}
        & \multirow{3}{*}{sign\_2p\_bes-shrink} & 0.05 & 0.531 & 0.593 & 0.602 \\
        & & 0.1 & \textbf{0.598} & \textbf{0.630} & \textbf{0.629} \\
        & & 0.2 & 0.531 & 0.575 & 0.615 \\
        \cmidrule{2-6}
        & \multirow{3}{*}{sign\_2p\_gs} & 0.05 & 0.583 & 0.595 & 0.593 \\
        & & 0.1 & \textbf{0.585} & \textbf{0.610} & \textbf{0.635} \\
        & & 0.2 & 0.534 & 0.564 & 0.617 \\
        
        \midrule
        \multirow{12}{*}{zaleplon\_mpo} & \multirow{3}{*}{adam\_2p\_bes-shrink} & 0.1 & \textbf{0.386} & 0.445 & 0.449 \\
        & & 0.2 & 0.208 & \textbf{0.447} & \textbf{0.483} \\
        & & 0.3 & 0.151 & 0.376 & 0.470 \\
        \cmidrule{2-6}
        & \multirow{3}{*}{adam\_2p\_gs} & 0.1 & \textbf{0.455} & \textbf{0.465} & 0.472 \\
        & & 0.2 & 0.374 & 0.453 & \textbf{0.491} \\
        & & 0.3 & 0.321 & 0.425 & 0.483 \\
        \cmidrule{2-6}
        & \multirow{3}{*}{sign\_2p\_bes-shrink} & 0.05 & 0.382 & 0.398 & 0.429 \\
        & & 0.1 & \textbf{0.478} & \textbf{0.485} & 0.477 \\
        & & 0.2 & 0.410 & 0.445 & \textbf{0.485} \\
        \cmidrule{2-6}
        & \multirow{3}{*}{sign\_2p\_gs} & 0.05 & 0.436 & 0.442 & 0.429 \\
        & & 0.1 & \textbf{0.460} & \textbf{0.487} & \textbf{0.488} \\
        & & 0.2 & 0.399 & 0.441 & 0.483 \\
        
        \midrule
        \multirow{18}{*}{deco\_hop} & \multirow{3}{*}{adam\_2p\_bes-shrink} & 0.1 & 0.544 & 0.564 & 0.605 \\
        & & 0.2 & 0.578 & \textbf{0.636} & 0.722 \\
        & & 0.3 & \textbf{0.585} & 0.628 & \textbf{0.738} \\
        \cmidrule{2-6}
        & \multirow{3}{*}{adam\_2p\_gs} & 0.1 & 0.564 & 0.585 & 0.603 \\
        & & 0.2 & 0.603 & 0.638 & 0.735 \\
        & & 0.3 & \textbf{0.612} & \textbf{0.669} & \textbf{0.741} \\
        \cmidrule{2-6}
        & \multirow{2}{*}{grad\_2p\_bes-shrink} & 50.0 & 0.480 & 0.587 & 0.666 \\
        & & 70.0 & \textbf{0.508} & \textbf{0.634} & \textbf{0.739} \\
        \cmidrule{2-6}
        & \multirow{4}{*}{grad\_2p\_gs} & 2.0 & 0.554 & 0.544 & 0.543 \\
        & & 5.0 & \textbf{0.608} & 0.597 & 0.584 \\
        & & 7.0 & 0.596 & 0.681 & 0.653 \\
        & & 10.0 & 0.592 & \textbf{0.696} & \textbf{0.688} \\
        \cmidrule{2-6}
        & \multirow{3}{*}{sign\_2p\_bes-shrink} & 0.1 & 0.645 & 0.709 & 0.784 \\
        & & 0.2 & \textbf{0.663} & \textbf{0.748} & \textbf{0.860} \\
        & & 0.3 & 0.613 & 0.670 & 0.746 \\
        \cmidrule{2-6}
        & \multirow{3}{*}{sign\_2p\_gs} & 0.1 & \textbf{0.676} & 0.763 & 0.762 \\
        & & 0.2 & 0.657 & \textbf{0.783} & \textbf{0.865} \\
        & & 0.3 & 0.616 & 0.621 & 0.763 \\
        \bottomrule
    \end{tabular}
\end{table}

\begin{table}[ht]
    \caption{Tuning of learning rate $\alpha$ for QED task. Scores correspond to the average of the success rates of optimizing 100 molecules after 1, 5, and 20 random restarts with $T=20$ iterations per restart.}
    \label{tab:tuning-qed}
    \centering
    \begin{tabular}{l|c|c|c|c}
        \toprule
        Methods & Learning rate $\alpha$ & $Q=30$ & $Q=50$ & $Q=100$ \\
        \midrule
        \multirow{3}{*}{adam\_1p\_bes-shrink} & 0.05 & 0.007 & 0.013 & 0.050 \\
        & 0.1 & 0.193 & 0.193 & 0.290 \\
        & 0.2 & \textbf{0.287} & \textbf{0.277} & \textbf{0.403} \\
        \midrule
        \multirow{3}{*}{adam\_1p\_gs} & 0.05 & 0.003 & 0.010 & 0.030 \\
        & 0.1 & 0.180 & 0.20 & 0.237 \\
        & 0.2 & \textbf{0.317} & \textbf{0.290} & \textbf{0.350} \\
        \midrule
        \multirow{3}{*}{adam\_2p\_bes-shrink} & 0.05 & 0.187 & 0.363 & 0.497 \\
        & 0.1 & 0.443 & 0.577 & 0.730 \\
        & 0.2 & \textbf{0.623} & \textbf{0.730} & \textbf{0.777} \\
        \midrule
        \multirow{3}{*}{adam\_2p\_gs} & 0.05 & 0.337 & 0.400 & 0.507 \\
        & 0.1 & 0.627 & 0.707 & 0.777 \\
        & 0.2 & \textbf{0.723} & \textbf{0.733} & \textbf{0.787} \\
        \midrule
        \multirow{2}{*}{grad\_1p\_bes-shrink} & 0.5 & 0.053 & 0.070 & 0.123 \\
        & 1.5 & \textbf{0.440} & \textbf{0.507} & \textbf{0.590} \\
        \midrule
        \multirow{3}{*}{grad\_1p\_gs} & 0.1 & 0.420 & 0.243 & 0.100 \\
        & 0.2 & \textbf{0.437} & \textbf{0.497} & \textbf{0.453} \\
        & 0.5 & 0.120 & 0.220 & 0.323 \\
        \midrule
        \multirow{3}{*}{grad\_2p\_bes-shrink} & 10.0 & 0.137 & 0.357 & 0.767 \\
        & 20.0 & \textbf{0.283} & \textbf{0.633} & \textbf{0.840} \\
        \midrule
        \multirow{4}{*}{grad\_2p\_gs} & 0.2 & 0.103 & 0.090 & 0.087 \\
        & 0.5 & 0.350 & 0.323 & 0.260 \\
        & 1.5 & \textbf{0.750} & 0.770 & 0.743 \\
        & 2.0 & 0.697 & \textbf{0.793} & \textbf{0.780} \\
        \midrule
        \multirow{3}{*}{sign\_1p\_bes-shrink} & 0.05 & 0.010 & 0.027 & 0.040 \\
        & 0.1 & 0.257 & 0.287 & 0.317 \\
        & 0.2 & \textbf{0.453} & \textbf{0.490} & \textbf{0.503} \\
        \midrule
        \multirow{3}{*}{sign\_1p\_gs} & 0.05 & 0.000 & 0.017 & 0.010 \\
        & 0.1 & 0.173 & 0.270 & 0.287 \\
        & 0.2 & \textbf{0.447} & \textbf{0.480} & \textbf{0.500} \\
        \midrule
        \multirow{3}{*}{sign\_2p\_bes-shrink} & 0.05 & 0.177 & 0.360 & 0.510 \\
        & 0.1 & 0.497 & 0.723 & 0.810 \\
        & 0.2 & \textbf{0.670} & \textbf{0.833} & \textbf{0.890} \\
        \midrule
        \multirow{3}{*}{sign\_2p\_gs} & 0.05 & 0.293 & 0.373 & 0.483 \\
        & 0.1 & 0.677 & 0.730 & 0.813 \\
        & 0.2 & \textbf{0.777} & \textbf{0.800} & \textbf{0.860} \\
        \bottomrule
    \end{tabular}
\end{table}

\section{Licenses}
All test sets of molecules were originally extracted from the ZINC database \citep{sterling2015zinc} which is free for use by anyone.

\end{document}